\documentclass[pra,aps, showpacs, twocolumn, a4paper,tightenlines,balancelastpage,superscriptaddress]{revtex4}
\usepackage{amsmath,amsfonts,amssymb,mathrsfs,bm,verbatim,psfrag,times}
\usepackage{graphicx,graphics,color,epsfig}
\usepackage{flafter,balance}
\usepackage{subeqnarray}
\usepackage{indentfirst,units}
\newcommand{\ket}[1]{\left|{#1}\right\rangle}
\newcommand{\bra}[1]{\left\langle{#1}\right|}

\begin{document}

\title{Probing multipartite entanglement in a coupled Jaynes-Cummings system}
\author{Peng \surname{Xue}}
\affiliation{Department of Physics, Southeast University, Nanjing 211189, P. R. China}
\affiliation{Institute for Quantum Information Science, University of Calgary, Alberta T2N 1N4, Canada}
\author{Zbigniew \surname{Ficek}}
\affiliation{The National Centre for Mathematics and Physics, KACST, P.O. Box 6086, Riyadh 11442, Saudi Arabia}
\author{Barry C.\ Sanders}
\affiliation{Institute for Quantum Information Science, University of Calgary, Alberta T2N 1N4, Canada}

\date{\today}

\begin{abstract}
We show how to probe multipartite entanglement in $N$ coupled
Jaynes-Cummings cells where the degrees of freedom are the
electronic energies of each of the $N$ atoms in separate single-mode
cavities plus the $N$ single-mode fields themselves. Specifically we
propose probing the combined system as though it is a dielectric
medium. The spectral properties and transition rates directly reveal
multipartite entanglement signatures. It is found that the Hilbert
space of the $N$ cell system can be confined to  the totally
symmetric subspace of two states only that are maximally-entangled W
states with $2N$ degrees of freedom.
\end{abstract}

\maketitle

\section{introduction}

A single two-level atom coupled to a single cavity, or resonator,
mode has been studied intensively since the introduction of this
`Jaynes-Cummings (JC) Model' independently by Jaynes and
Cummings~\cite{JC63} and by Paul~\cite{Pau63} in 1963. Quantum-field
effects such as periodic spontaneous collapse and
revival~\cite{ENS80} are now studied and observed in many systems
and as a multitude of manifestations~\cite{CKS96}. Coupled JC systems
have been proposed as a basis for quantum networks~\cite{PK88} and could behave as a novel
condensed-matter system if enough JC systems can be coupled together~\cite{HBP06}.

In a preliminary study we developed theoretical tools for
calculating the spectrum, stationary states and dielectric
susceptibility~\cite{XFS11}. Here we use and extend those tools,
especially to include the nontrivial open-system effects of
spontaneous emission and cavity losses. In particular, we consider the problem
of how to probe and characterize such systems experimentally.
Two quite different approaches are evident.
One approach is to probe each component in microscopic paradigm,
namely drive and detect the various atoms and cavity modes.
Another approach, which we favour, is to treat the coupled JC system as a `black-box' model
and probe it as a single unit following a macroscopic paradigm.

Our concept is to regard the coupled JC system as a dielectric medium whose susceptibility carries a signature of the peculiarities of the coupled JC system. Diagonalizing the Hamiltonian of the coupled JC system, we find entangled states and their energies. Using the Fermi's golden rule, we calculate the transition rates between different manifolds of the energy states of the system. In this paper, we restrict the calculation to transitions from the single-excitation states to the ground state of the system.
In particular we show that the dielectric susceptibility of the coupled-JC medium reveals,
by probe-field spectroscopy,
quadripartite entanglement of the system comprising mutually coupled atoms and cavity modes.
Based on our theoretical framework for JC systems mutually coupled
by overlapping extra-cavity longitudinal fields,
we can calculate stationary states for the coupled system, energy spectrum, and dielectric susceptibility.

The paper is organized as follows. In Sec.~\ref{sec:composite} we give a qualitative discussion and a detailed calculation of radiative properties of a single JC cell. Section~\ref{sec3} is devoted to the discussion of the entangled and radiative properties of two coupled JC cells. We derive single-excitation states of the system and show that they are the W~state class corresponding to a superposition state of single excitations amongst each of the two atoms and two modes. We then find under which conditions the states reduce to the maximally entangled four-qubit W states, and how to quantify the degree of entanglement of the states using probe-field spectroscopy methods.
A generalization of the calculation to an arbitrary number of mutually coupled JC cells is presented in Sec.~\ref{sec4}. Finally, in Sec.~\ref{sec5}, we summarize our results.

\section{Entanglement and radiative properties of the single JC cell}
\label{sec:composite}

The JC cell (as we refer to a single JC system) is a composite atom cavity-mode system that radiates via two distinct channels.
The closed coupled JC system is characterized by just two parameters,
namely the atom-cavity coupling rate~$g$ and the inter-cavity coherent hopping rate~$\kappa$.
One radiative channel arises due to coupling of the atom to the cavity side modes
(free-space modes) causing spontaneous decay of the atomic
excitation with an inhibited spontaneous emission rate~$\gamma_\text{a}$.
The inhibition refers to the fact that some free radiative modes are suppressed by cavity confinement of the atom.
The other radiative decay mode is due to losses from one or both mirrors of the optical cavity at rate  $\gamma_\text{c}$. We assume independence between these two radiative channels.

A single JC cell comprises a single cavity field mode of frequency
$\omega_\text{c}$ and a two-level atom with energy states $\ket{g_i}$
and~$\ket{e_i}$ and corresponding energy difference
\begin{equation}
    \hbar\omega_\text{a}=E_e-E_g.
\end{equation}
The atom is coupled to the cavity mode with coupling constant~$g$, which we
choose to be real with no loss of generality. We assume that the
atomic transition frequency is detuned from the cavity frequency
by~$\Delta =\omega_\text{a}-\omega_\text{c}$.

In the rotating-wave approximation, and using a system of units in
which $\hbar\equiv 1$, the Hamiltonian of the single JC system~is
\begin{equation}
\hat{H}^\text{JC} =\omega_\text{c}\left(\hat{a}^\dagger \hat{a}
+\frac{1}{2}\right)+\frac{1}{2}\omega_\text{a}\hat{\sigma}_{z}
        +g\left(\hat{a}^\dagger\hat{\sigma}_{-} +\hat{a}\hat{\sigma}_{+}\right) ,
\end{equation}
with~$\hat{a}^{\dagger}$ and $\hat{a}$ creation
and annihilation operators of the cavity mode, respectively, and
$\hat{\sigma}_{\pm}$ and $\hat{\sigma}_{z}$, with
$\text{spec}(\hat\sigma_{z}^{i})=\pm1$, the spin operators for the atom.

\subsection{Energy spectrum}

The JC energy spectrum comprises a ground state of energy~$\omega_0=-\Delta/2$
and a ladder of doublets of energies
\begin{equation}
    \omega_{\pm n} =n\omega_\text{c}\pm\sqrt{ng^2+\frac{1}{4}\Delta^2},\quad n=1,2,\ldots
\end{equation}
The corresponding stationary energy states are the ground
state~$\ket{0g}$ and the excited state doublets
\begin{align}
    \ket{+,n} =& \cos\theta_{n}\ket{e,n-1} +i\sin\theta_{n}\ket{g,n} ,\nonumber\\
    \ket{-,n} =& -\sin\theta_{n}\ket{e,n-1} +i\cos\theta_{n}\ket{g,n} ,\label{eq:ket+-n}
\end{align}
with
\begin{equation}
\label{eq:theta_n}
    2\theta_n = \tan^{-1}\left(\frac{2g\sqrt{n}}{\Delta}\right).
\end{equation}
Note that the doublets are in a pure, bipartite entangled state of the atom and the field. This entanglement is maximal (for all the doublets) only when $\Delta =0$.
\begin{figure}[ht]
\includegraphics[width=4.1cm]{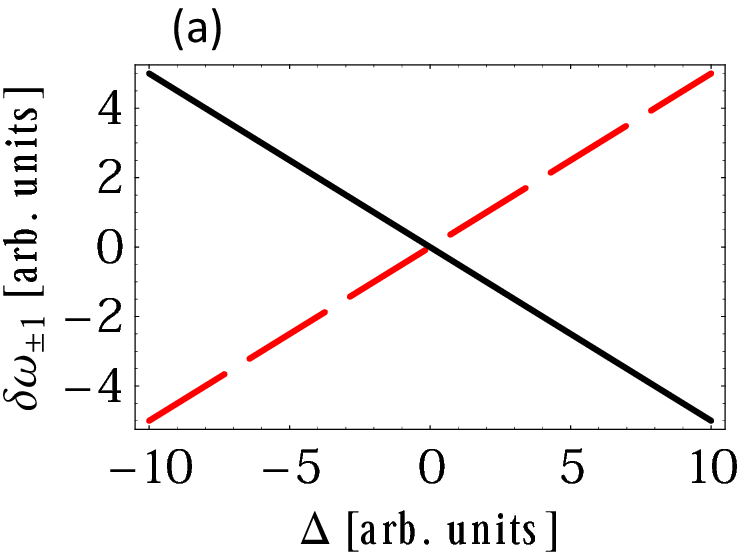}
\includegraphics[width=4.1cm]{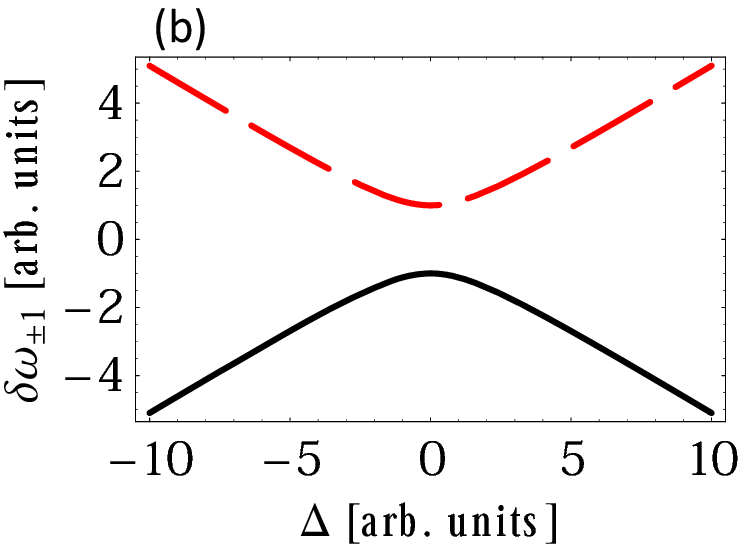}
\caption{(Color online) Variation of the energy difference
$\delta\omega_{\pm 1} =\omega_{\pm 1} - \omega_{\text{c}}$ of the
$n=1$ eigenstates with the detuning $\Delta$ (in arbitrary units);
$\delta\omega_{-1}$ (black solid line) and $\delta\omega_{+1}$ (red
dashed line) for a single JC cell with: (a)~$g=0$ and~(b)~$g=1$.}
\label{fig:spectra1}
\end{figure}

Figure~\ref{fig:spectra1} shows a variation of the energies $\omega_{\pm n}$ of the $n=1$ doublet $\ket{\pm,1}$ with the detuning $\Delta$ around the unperturbed energy $\omega_{\text{c}}$. Notice the level crossing in the absence of the coupling $g$ at $\Delta =0$, and the appearance of the familiar avoided crossing effect when $g\neq 0$.

In the following we explore the radiative properties of this single JC cell, in particular looking for signatures of this atom-field entanglement in the transition rates between states with different total number of excitations. Specifically, we search for conditions that reveal whether the entanglement of the eigenstates (\ref{eq:ket+-n}) is maximal or not.

\subsection{Transition rates}

As we have seen, each doublet has some fixed number of quanta shared
between the atom's electronic state and the cavity mode field state.
We employ the integer~$\nu$ to designate the number of quanta in the
system. The ground state corresponds to $\nu=0$, which means that
the electronic state and field state are both in the lowest level.
The first doublet has $\nu=1$ quantum, the second doublet has
$\nu=2$ quanta, and so on.

Consider first the case $n=1$ with transitions
from the single-excitation states $\ket{\pm,1}$, with splitting due to the
atom-field interaction, to the ground state, which is a product
state not affected by the atom-field interaction.
The transition rates from the excited states to the ground state are
given by Fermi's golden rule~\cite{ct92}
\begin{align}
\label{eqg}
    \Gamma_{\pm,1}=\gamma_\text{a}\left|\bra{1,\pm}\hat{\sigma}_{+}\ket{g,0}\right|^2
        +\gamma_\text{c}\left|\bra{1,\pm}\hat{a}^{\dag}\ket{g,0}\right|^2 .
\end{align}
These transition rates are a sum of transitions
caused by spontaneous emission from the atom, with rate~$\gamma_\text{a}$,
and by damping of the cavity mode, with rate
$\gamma_\text{c}$.
Consequently the coefficient~$\gamma_\text{a}$ quantifies the amount that the atomic spontaneous emission
contributes to the transition probability.
Similarly, $\gamma_\text{c}$ quantifies how much the cavity losses  contribute to the transition probability.

From Eq.~(\ref{eq:ket+-n}) we readily calculate the transition dipole
moments between the states $\ket{\pm,1}$ and $\ket{g,0}$ whence
we obtain the transition probabilities
\begin{align}
\Gamma_{+,1} &= \gamma_\text{c} +\left(\gamma_\text{a}-\gamma_\text{c}\right)\cos^2\theta_{1},\nonumber\\
\Gamma_{-,1} &= \gamma_\text{a} -\left(\gamma_\text{a}-\gamma_\text{c}\right)\cos^2\theta_{1} .\label{eq:Gamma_+-1}
\end{align}
A number of interesting properties are immediately evident from the
expressions for these probabilities. For example, in the absence of
either atomic spontaneous emission $\left(\gamma_\text{a}=0\right)$
or cavity dissipation $\left(\gamma_\text{c}=0\right)$, that is when
only a single dissipation channel is present in the system, the
transition probabilities depend on the nature of the states from
which they originate with this nature given by the
parameter~$\theta_1$ appear in Eqs.~(\ref{eq:ket+-n}) and (\ref{eq:theta_n}).

This dependence on ~$\theta_1$ causes the two transition
probabilities to be mutually correlated. For example, an
increase of~$\Gamma_{+,1}$ implies a decrease of $\Gamma_{-,1}$ and
vice versa. The probabilities become independent of each other and
their magnitudes equalize only when the states become maximally
entangled. Thus, we may infer from the transition probabilities to what degree
the states are entangled. However, this conclusion is based on a
simplified model of the system involving only a single decay
channel, which is incompatible with what one encounters
in practice. A practical JC system radiates through both dissipation
channels.

We distinguish here two parameter regimes with a qualitatively
different behavior for the transition probabilities. These two
regimes are distinguished according to the magnitude of
$\gamma_\text{c}$ relative to~$\gamma_\text{a}$. When the
dissipation rates are equal,
$\gamma_\text{a}=\gamma_\text{c}\equiv\gamma$,
Eq.~(\ref{eq:Gamma_+-1}) yields that the probabilities are
independent of each other and have equal magnitudes,
$\Gamma_{+,1}=\Gamma_{-,1}=2\gamma$, regardless of the state of the
system (i.e.\ $\theta_1$). In other words, the transition
probabilities tell us nothing about the nature of the states
involved. In physical terms, this is a consequence of the fact that,
for~$\gamma_\text{a}=\gamma_\text{c}$, one cannot distinguish which
dissipation channel is used by the quantum leaving the JC system
regardless of the initial state. Therefore, determining the
entanglement of the states~$\ket{\pm,1}$ is not  possible from
transition rates if~$\gamma_\text{a}=\gamma_\text{c}$ so of course
the set-up must then avoid this condition for entanglement to be
measurable.

For~$\gamma_\text{a}\neq \gamma_\text{c}$, the transition rates
depend explicitly on the amplitudes of the states involved thereby
enabling the degree of entanglement to be discerned from the
transition rates, for example by measuring the difference or the
ratio between $\Gamma_{+,1}$ and $\Gamma_{-,1}$. From
Eq.~(\ref{eq:Gamma_+-1}) we see that, for non-maximally entangled
states $\left(\cos^2\theta_{1}\neq 1/2\right)$, equality between
$\Gamma_{+,1}$ and $\Gamma_{-,1}$ cannot be achieved. However, when
$\cos^2\theta_{1}=1/2$, which corresponds to the case of maximally
entangled states, $\Gamma_{+,1}$ and $\Gamma_{-,1}$ have the same
magnitude. Equality between the transition probabilities with
$\gamma_\text{a}\neq \gamma_\text{c}$ can only occur for maximally
entangled states.

We now consider the case $n\geq 2$. In this case, transitions occur
between two neighboring doublets of entangled states ($n=2$ and
$n=1$). Notice that the transitions occur between states of
different degree of entanglement, and there are two possible
transition channels from each state of the upper doublet to states
of the doublet below. Transitions from the $i^\text{th}$ to the
$j^\text{th}$ state of the neighboring~$n$ and~$n-1$ doublets occur
with rate
\begin{align}
    \Gamma_{i,jn} = \gamma_\text{a}\!\left|\bra{n,i}\!\hat{\sigma}^{+}\ket{j,n-1}\right|^2 + \gamma_\text{c}\!\left|\bra{n,i}\!a^{\dag}\!\ket{j,n-1}\right|^2 .
\label{eq:Gamma_i,jn}
\end{align}
Using Eqs.~(\ref{eq:ket+-n}) and (\ref{eq:Gamma_i,jn}), we readily find that the
transitions occur with probabilities
\begin{align}
\label{eq:Gamma_+pmn}
    \Gamma_{+,\pm n} &= \frac{1}{2}\left[\left(\gamma_\text{a}-\gamma_\text{c}\right)\mp\left(\gamma_\text{a}
        +\gamma_\text{c}\right)\cos2\theta_{n-1}\right]\cos^2\theta_{n} \nonumber\\
        &+\frac{1}{2}n\gamma_\text{c}\left(1 \pm\cos2\theta_{n}\cos2\theta_{n-1}\right)  \nonumber\\
        &\pm\frac{1}{2}\gamma_\text{c}\sqrt{n(n-1)}\sin2\theta_{n}\sin2\theta_{n-1} ,\nonumber\\
        \Gamma_{-,\pm n} &= \frac{1}{2}\left[\left(\gamma_\text{a}-\gamma_\text{c}\right)\mp\left(\gamma_\text{a}
        +\gamma_\text{c}\right)\cos2\theta_{n-1}\right]\sin^2\theta_{n} \nonumber\\
        &+\frac{1}{2}n\gamma_\text{c}\left(1 \mp \cos2\theta_{n}\cos2\theta_{n-1}\right)  \nonumber\\
        &\mp\frac{1}{2}\gamma_\text{c}\sqrt{n(n-1)}\sin2\theta_{n}\sin2\theta_{n-1} ,
\end{align}
thereby yielding
\begin{align}
\label{eq:Gamma_+n}
    \Gamma_{+,n} &= \sum_{j=\pm}\Gamma_{+,jn} = n\gamma_\text{c} + \left(\gamma_\text{a}-\gamma_\text{c}\right)\cos^2\theta_{n} ,\nonumber\\
    \Gamma_{-,n} &= \sum_{j=\pm}\Gamma_{-,jn} = n\gamma_\text{c} + \left(\gamma_\text{a}-\gamma_\text{c}\right)\sin^2\theta_{n} ,
\end{align}
for the total transition probabilities from the states~$\ket{\pm,n}$.

Although the transitions occur between states of
different degrees of entanglement, the total transition
probabilities are determined~\emph{only} by the amplitudes of
the states of the upper doublet. The states of the lower doublet do
not become involved. The properties of the total probabilities are
essentially similar to those discussed above for the transitions from
the~$n=1$ doublet to the ground state. Indeed, for $n=1$ the
probabilities~(\ref{eq:Gamma_+n}) reduce to Eq.~(\ref{eq:Gamma_+-1}),
and unequal damping rates, $\gamma_\text{a}\neq
\gamma_\text{c}$, ensures the dependence of the transition
probabilities on the degree of entanglement.

Thus, transition rates of a combined quantum
system depend strongly on how the subsystems decay rather than on
the nature of states involved. The presence of the maximally
entangled states in a JC cell could be observed in principle by
measuring the transition rates between energy levels of the system,
subject to~$\gamma_\text{a}\neq\gamma_\text{c}$.

\subsection{Absorption spectrum of a weak probe beam}

There remains the question how these properties of the transition
rates might be exhibited experimentally. Probe-beam spectroscopy
methods should be able to test these properties. When a JC system is
irradiated by a weak probe beam with a frequency that is close to
resonance, the rate the probe beam is absorbed is proportional to
the transition rates between the energy states of the system.

We consider net absorption by the system of radiation from a tunable
beam probing the system. The probe beam intensity is assumed to be
sufficiently weak such that the field is treated to only first order
in its amplitude so that it does not appreciably perturb the system.
The absorption spectrum is given by the imaginary part of the linear
susceptibility~$\chi ^{(1)}_{0}\left(\omega_{p}\right)$ of the
system~\cite{kubo,mol72,szym95}
\begin{align}
    \text{Im}\left[\chi ^{(1)}_{0}\left(\omega_{p}\right)\right]
    = \sum_{i,n}\frac{\gamma\Gamma_{i,n}}
    {\left(\omega_{in} -\omega_{p}\right)^2 + \gamma^{2}}  ,\label{equ8}
\end{align}
where $\omega_{p}$ is the frequency of a probe beam, $\Gamma_{i,n}$ is
the total transition rate from the state $i\, (i=\pm)$ of the manifold $n$,~$\gamma$ describes width of a given transition. In writing expression~(\ref{equ8}), we have used the fact that
transitions from the excitation states to states of the manifold below might not be
purely radiative~\cite{BR81,ct92}, i.e. $\gamma\neq \Gamma_{i,n}/2$, and the system was in the $n-1$ manifold before the probe was applied. To ensure that the spectral lines are well resolved, we assume that the
transitions do not overlap. This is achieved assuming that the coupling strength $g$ is much larger than the transition rates~$\Gamma_{i,n}$.
\begin{figure}[ht]
    \includegraphics[width=6.1cm]{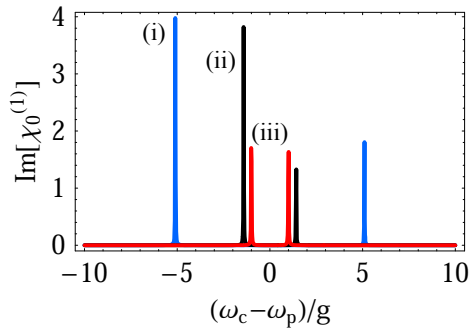}
    \caption{(Color online) The absorption spectra ${\rm Im}\left[\chi^{(1)}_{0}(\omega_\text{p})\right]$ of a
    probe beam monitoring a single JC cell for $\gamma_\text{a}/g=0.05$,  $\gamma_\text{c}/g=0.02$, $\gamma/g=0.01$,
    and different detunings: $\Delta/g=0$ (red line (iii)), $\Delta/g=2$ (black line (ii)), and $\Delta/g=10$ (blue line (i)).}
\label{fig:ab1cell}
\end{figure}

Figure~\ref{fig:ab1cell} shows absorption spectra of a probe beam tuned in vicinity of the transition frequencies from the $n=1$ doublet to the ground state of a single JC cell.
We see that as long as $\Delta\neq 0$, the spectrum of the single cell is markedly asymmetric. This feature is associated with the fact that at $\Delta\neq 0$ the transitions from the $n=1$ doublet to the ground state occur with different rates, $\Gamma_{+,1}\neq \Gamma_{-,1}$. The spectrum becomes symmetric at resonance, where $\Delta =0$. The symmetric spectrum is associated with the fact that at the resonance, $\Gamma_{+,1} =\Gamma_{-,1}$. As predicted by Eq.~(\ref{eq:Gamma_+-1}), the equality of the total transition rates takes place only for the maximally entangled states. Therefore, the symmetry of the spectrum can be regarded as an indication of the presence of maximally entangled states.

\section{Two mutually coupled JC cells}\label{sec3}

We now consider two neighboring JC cells coupled via overlapping evanescent waves of the cavity modes.
The coupling results in an coherent hopping rate~$\kappa$ between cavities. We
designate~$\hat{a}_i$ and~$\hat{\sigma}_i$ as the field annihilation
operator and atomic electron energy lowering operator for the
$i^\text{th}$ JC cell, and the double JC cell Hamiltonian is~\cite{sb99,nr07,oi08,PMM07,XFS11}
\begin{equation}
    \hat{H} = \hat{H}_{1}^{\text{JC}} +\hat{H}_{2}^{\text{JC}} -\kappa\left(\hat{a}_{1}^{\dagger}\hat{a}_{2}
        +\hat{a}_{1}\hat{a}_{2}^{\dagger}\right) =\bigoplus_\nu\hat{H}^{(\nu)} ,\label{djch}
\end{equation}
for~$\nu$ denoting the total number of quanta shared in the two-cavity system.

The space of the system for any $\nu$ is spanned by product states $\{\ket{n_1,c_1,n_2,c_2}\}$ with $n_i$ a label for the number of photons in the $i^\text{th}$ cavity
and $c_i$ the label for whether the state is in the excited ($\ket{e}$) or ground ($\ket{g}$) state.
For example, the ground state corresponds to $\nu=0$ with just one dimension hence contains
one basis element~$\{\ket{0g0g}\}$.
The single-excitation $(\nu=1)$ basis comprises four states $\{\ket{0e0g},\ket{1g0g},\ket{0g1g},\ket{0g0e}\}$.

In the following, we limit ourselves to the subspace $\nu =1$ that only a single excitation is present in the system. The $\nu=1$ case is especially interesting as it corresponds to four qubits:
two two-level atoms and two field modes with superpositions of vacuum and single-photon states.
Also the $\nu=1$ case can be solved in closed form yielding simple expressions~\cite{XFS11}.

\subsection{Entangled four-qubit states}

A diagonalization of the Hamiltonians $\hat{H}^{(\nu)}$ for $\nu=0,1$ yields the energy
spectra~$\omega^{(0)}=-\Delta$, for the ground state, and
\begin{equation}
    \omega^{(1)}_{\epsilon\varepsilon}
        =\omega_\text{c}-\frac{1}{2}(\Delta+\epsilon\kappa)+\varepsilon\sqrt{g^2+\frac{1}{4}\left(\Delta+\epsilon\kappa\right)^2}
\end{equation}
for the single-excitation, with~$\epsilon=\pm$ and $\varepsilon=\pm$. In the case of independent cells, $\kappa=0$, and then the upper spectral value is degenerate
\begin{equation}
    \omega^{(1)}_{\epsilon\varepsilon}|_{\kappa=0} =\omega^{\text{JC}(0)} +\omega_\varepsilon^{\text{JC}(1)} ,
\end{equation}
and the spectrum corresponds to what is expected for two independent JC systems, with~$\epsilon$
being irrelevant.

On the other hand, for $g=0=\Delta$, the $\nu=0$ spectrum is characterized by $\omega^{(0)}=0$, whereas the $\nu=1$ is described by  $\omega^{(1)}_{\epsilon\varepsilon}|_{g=0=\Delta}$, which is  the doubly-degenerate~$\omega_\text{c}$ or $\omega_\text{c}+\varepsilon\kappa$ values as expected for coupled
harmonic oscillators. Therefore, a large inter-cavity hopping rate
will push the coupled JC cells to behaving closely like coupled
harmonic oscillators with atom-cavity perturbations. For small~$\kappa$
\begin{equation}
    \omega^{(1)}_{\epsilon\varepsilon}\approx\omega_\text{c} +\varepsilon g-\frac{\epsilon\kappa}{2}
    +\varepsilon\frac{\kappa^2}{8g} ,
\end{equation}
with shift $\pm g$ due to vacuum Rabi splitting, $\epsilon\kappa/2$
the normal-mode splitting due to inter-cavity coupling and
$\varepsilon\kappa^{2}/8g$ a frequency pulling effect similar to the ac Stark shift.
\begin{figure}[ht]
       \includegraphics[width=4.1cm]{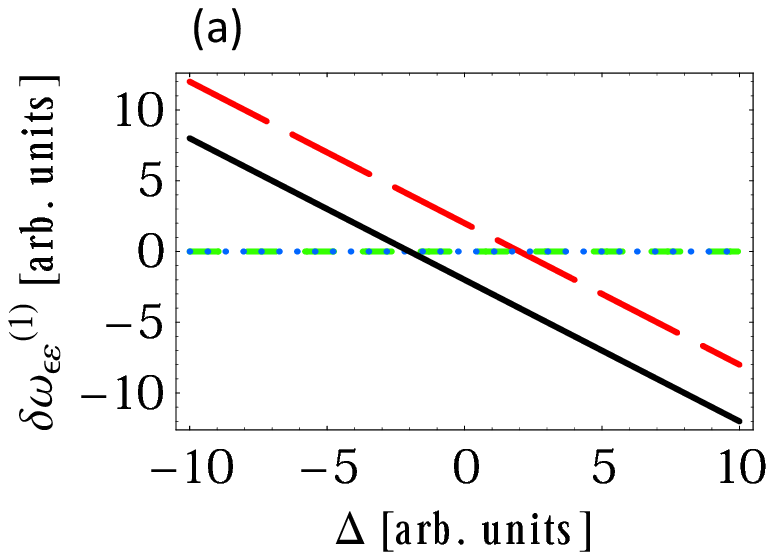}
       \includegraphics[width=4.1cm]{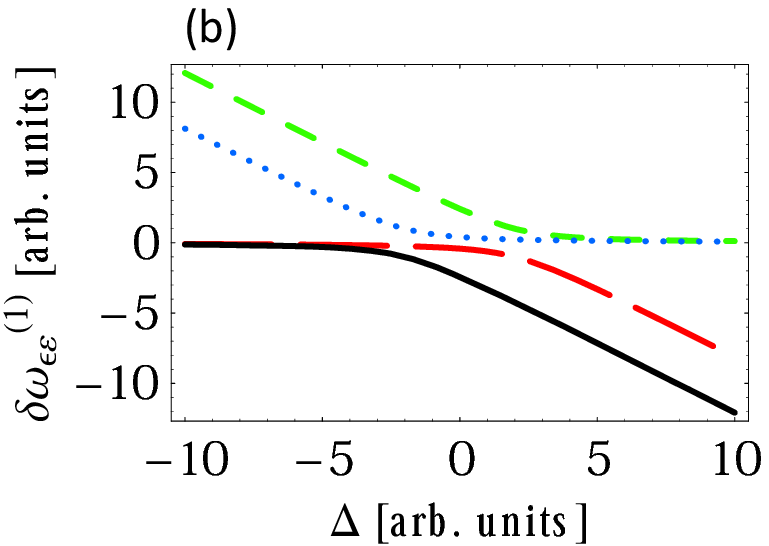}
    \caption{(Color online) The dependence of the energy difference
    $\delta\omega^{(1)}_{\epsilon\varepsilon} = \omega^{(1)}_{\epsilon\varepsilon}-\omega_{\text{c}}$
    on the detuning $\Delta$ (in arbitrary units) of the~$\nu =1$ states,~$\ket{+,-}$ (black solid line),
    $\ket{-,-}$ (red long dashed  line), $\ket{-,+}$ (green short dashed line), $\ket{+,+}$ (blue dotted line),
    for two coupled JC cells with (a)~$g=0, \kappa=2$, and (b)~$g=1, \kappa=2$.}
    \label{fig:spectra2}
\end{figure}

Figure~\ref{fig:spectra2} shows the variation of the energies $\omega^{(1)}_{\epsilon\varepsilon}$ of the single excitation $\nu=1$ states with the detuning $\Delta$ around the unperturbed energy $\omega_{\text{c}}$. In the absence of the coupling $g$, there are two crossing points at $\Delta =\pm \kappa$. The coupling lifts the degeneracy resulting in four entangled states. The states are eigenstates of the Hamiltonian (\ref{djch}) corresponding to the energies $\omega^{(1)}_{\epsilon\varepsilon}$, and are of the form~\cite{XFS11}
\begin{equation}
\label{eq:states}
    \ket{\epsilon,\varepsilon} = u_{\epsilon,\varepsilon}\!\left(\ket{1g0g}\!-\!\epsilon\ket{0g1g}\right)\!
        +\!w_{\epsilon,\varepsilon}\!\left(\ket{0e0g}\!-\epsilon\!\ket{0g0e}\right) ,
\end{equation}
where
\begin{align}
\label{uw}
    u_{\epsilon,\varepsilon}
        &= \frac{-r_\epsilon+\varepsilon\sqrt{1+r_{\epsilon}^2}}{\sqrt{2+2\left(r_{\epsilon}-\varepsilon\sqrt{1+r_{\epsilon}^2}\right)^2}}  ,\nonumber\\
    w_{\epsilon,\varepsilon}
        &= \frac{1}{\sqrt{2+2\left(r_{\epsilon}-\varepsilon\sqrt{1+r_{\epsilon}^2}\right)^2}} ,
\end{align}
and
\begin{equation}
r_\epsilon = (\Delta+\epsilon\kappa)/(2g) .
\end{equation}
The double-JC states~(\ref{eq:states}) are in the W~state class
corresponding to a superposition state of single excitations amongst
each of the two atoms and two modes. These states are maximally
entangled only for $u_{\epsilon,\varepsilon} =w_{\epsilon,\varepsilon} =\nicefrac{1}{2}$,
which requires that $r_{\epsilon} =0$. The condition that
$r_{\epsilon} =0$ is met only if $\Delta=\varepsilon \kappa$.
Radiative properties of two coupled JC cells are studied in the next
section.

\subsection{Transition rates and their collective properties}

We can compute the transition probabilities~\cite{HF79,RFD95,BP07}
\begin{equation}
\Gamma_{\epsilon,\varepsilon} = \gamma_\text{a}|\!\bra{\epsilon,\varepsilon}\!\left(\hat{\sigma}_{1}^{+}
+\hat{\sigma}_2^{+}\right)\!\ket{0}\!|^2\!+\gamma_\text{c}|\!\bra{\epsilon,\varepsilon}\!(a_{1}^{\dag} + a_2^{\dag})\!\ket{0}\!|^{2}
\label{eq14}
\end{equation}
from the single excitation states $\ket{\epsilon,\varepsilon}$ to the ground state~$\ket{0{\text g}0\text{g}}\equiv \ket{0}$, which yields
\begin{equation}
    \Gamma_{\epsilon,\varepsilon} = \left(1-\epsilon\right)^2\left(\gamma_\text{a}
    \left|w_{\epsilon,\varepsilon}\right|^2
        +\gamma_\text{c}\left|u_{\epsilon,\varepsilon}\right|^2\right) .\label{eq15}
\end{equation}
The dependence of the transition rates on $\epsilon$ signals the
existence of superradiant $(\epsilon=-)$ and metastable
non-radiative $(\epsilon =+)$ states in the system~\cite{dic,FT02}.
Note that the equality of the superradiant rates, $\Gamma_{-,+}=\Gamma_{-,-}$,
is a signature of the maximal entanglement present in the
corresponding states. However,
this signature is not present for the other two states $|+,\varepsilon\rangle$,
as these do not imprint a signature on the radiation properties
of the system.

Notice that these collective radiative properties of the system are
independent of the ratio $\gamma_\text{a}/\gamma_\text{c}$ and of
the nature of the states. These decay processes are due to by losses
of the coupled cavity modes as well as by atomic spontaneous
emission. This latter feature is surprising because atoms are not
directly coupled to each other, and they behave independent of the
ratio $\kappa/g$. The existence of the metastable states implies
that the double JC system effectively behaves as a single collective
JC cell composed of a doublet consisting of four-qubit entangled
states $\ket{\epsilon,-}$.

The metastable states could be made radiatively active by breaking
the symmetry between the atoms and/or the cavity modes of {\it different}
JC cells, for example, by allowing the atoms and the
cavity modes to decay at different rates. It is easy to show that
when the atoms are damped with different rates, say
$\gamma_{\text{a}1}$ and $\gamma_\text{a2}$, and the cavity losses
occur with rates~$\gamma_{\text{c}1}$ and $\gamma_\text{c2}$,
respectively, then the transitions from the states
$\ket{\epsilon,\varepsilon}$ occur with probabilities
\begin{equation}
\label{eq:Gammaepsilonpm}
    \Gamma_{\epsilon,\varepsilon} = \left(\sqrt{\gamma_{\text{a}1}}
        -\epsilon\sqrt{\gamma_\text{a2}}\right)^2\!\left|w_{\epsilon,\varepsilon}\right|^2
            + \left(\sqrt{\gamma_{\text{c}1}} -\epsilon\sqrt{\gamma_\text{c2}}\right)^2
                \left|u_{\epsilon,\varepsilon}\right|^2.
\end{equation}
Clearly, the transition probabilities are different from zero
irrespective of $\epsilon$. However, the states retain their
collective character with the states $\epsilon =-$ still behaving
as a superradiant and the $\epsilon=+$ states now behaving as a subradiant state.

There are some similarities in the properties of the transition
probabilities of the double- and single-JC cell systems, in
particular, when the atomic spontaneous emission and cavity losses
in a JC cell occur with the same rate. For example, in the case of
$\gamma_{\text{a}1}=\gamma_{\text{c}1}$ and
$\gamma_\text{a2}=\gamma_\text{c2}$, the
probabilities~(\ref{eq:Gammaepsilonpm}) become independent of the
amplitudes of the states, which is the same property  encountered
for $\Gamma_{\pm 1}$ of the single cell.

However, there are interesting differences. In the double~JC system
the condition $\gamma_\text{ai}\neq \gamma_\text{ci}$ is necessary
but not sufficient for the dependence of the transition rates on the
amplitudes of the states. There is also a rather subtle condition of
the relation between the damping rates of different cells to be
satisfied. Even though the spontaneous emission and cavity losses in
a given~JC cell occur with different rates, $\gamma_\text{ai}\neq
\gamma_\text{ci}$, the transition probabilities still could be
independent of the amplitudes of the states. This happens when
$\gamma_{\text{a}1}= \gamma_\text{c2}$ and
$\gamma_\text{a2}=\gamma_{\text{c}1}$, for which
Eq.~(\ref{eq:Gammaepsilonpm}) reduces to
\begin{equation}
    \Gamma_{\epsilon,\varepsilon} = \frac{1}{2}\left(\sqrt{\gamma_{\text{a}1}}
    -\epsilon\sqrt{\gamma_{\text{c}1}}\right)^2  .\label{eq20}
\end{equation}

Evidently, the transition rates are independent of the amplitudes of
the states. The quantitative reason for this is that, even if the
cells are distinguished by different damping rates, they are
directly coupled to each other through the coupling~$\kappa$.
This coupling creates entangled states between the cells that in the case
of $\gamma_{\text{a}1}= \gamma_\text{c2}$ makes indistinguishable
through which channel, $\gamma_{\text{a}1}$ or $\gamma_\text{c2}$, a
photon leaving the system was emitted. This property is
characteristic of multi-cell JC systems and does not exist in a
single JC cell.
Therefore, under the condition that all the
damping rates are different, equality between the superradiant
(subradiant) rates signals maximal entanglement in the superradiant
(subradiant) states, $\ket{-,\varepsilon}$ ($\ket{+,\varepsilon}$), which
generalizes the results of the single JC cell.

\subsection{Absorption spectra}

We now consider absorption spectra of a weak radiation monitoring the transitions from the single-excitation
$(\nu =1)$ states to the ground state $\ket 0$ of two identical JC cells. To ensure that the probe field couples
exclusively to the single excitation states, we assume that the
transition do not overlap. This is achieved assuming that the
coupling strengths $g$ and $\kappa$ are much larger than the
transition rates $\Gamma_m$.
\begin{figure}[ht]
    \includegraphics[width=6.1cm]{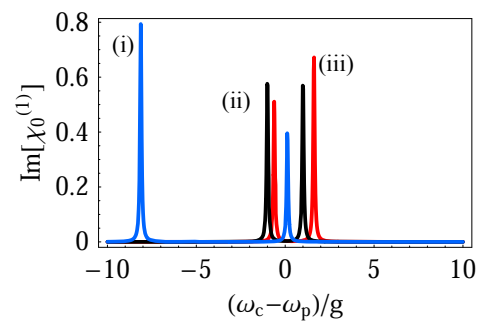}
    \caption{(Color online) The absorption spectra
    ${\rm Im}\left[\chi^{(1)}_{0}(\omega_\text{p})\right]$ of a probe beam monitoring two coupled
    identical JC cells, for $\gamma_\text{a}/g=0.05$,  $\gamma_\text{c}/g=0.02$, $\gamma/g=0.01$, $\kappa/g=2$,
    and different detunings: $\Delta/g=1$ (red line (iii)), $\Delta/g=2$ (black line (ii)), and $\Delta/g=10$ (blue line (i)).}
\label{fig:ab2cell}
\end{figure}

Figure~\ref{fig:ab2cell} shows absorption spectra for $\gamma_\text{a}\neq \gamma_\text{c}$ and different detunings $\Delta$. We see that as long as $\Delta\neq \pm\kappa$, the spectrum is always asymmetric.
The spectrum becomes symmetric when $\Delta=\kappa$. In this case the states $\ket{\pm,-}$ are maximally entangled states. Thus, similar to the case of a single JC cell, the symmetry of the spectrum can be regarded as an indication of the presence of maximally entangled states. Note that the symmetric spectrum is observed at non-zero detunings, $\Delta=\pm\kappa$, which is in contrast to the properties of the spectrum of a single cell, where the symmetric spectrum is observed only at resonance, where $\Delta=0$.
\begin{figure}[ht]
    \includegraphics[width=4.1cm]{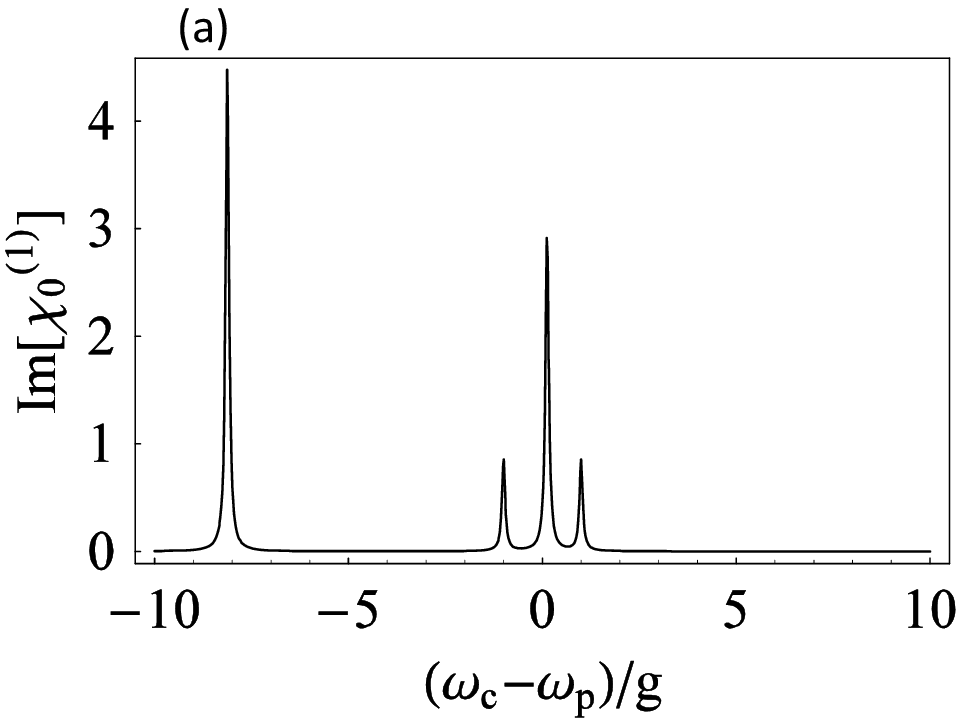}
    \includegraphics[width=4.1cm]{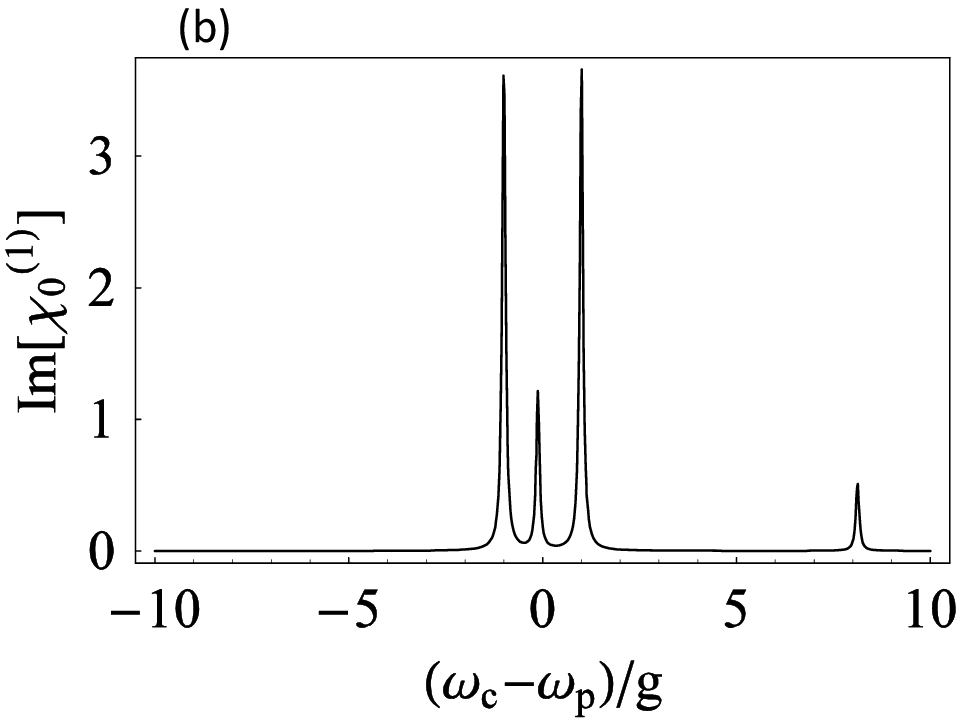}
    \caption{(Color online) The absorption spectra
    ${\rm Im}\left[\chi^{(1)}_{0}(\omega_\text{p})\right]$
    of a probe beam monitoring the system of two nonidentical JC cells for
    $(\gamma_\text{a1}/g,\gamma_\text{a2}/g) =(0.01,0.2)$,
$(\gamma_\text{c1}/g,\gamma_\text{c2}/g)=(0.2,0.05)$, $\kappa/g=4$, $\gamma/g=0.05$,
and different $\Delta$: (a)~$\Delta/g=-4$ and (b)~$\Delta/g=4$.}
\label{fig:4n}
\end{figure}

Figure~\ref{fig:4n} shows the absorption spectrum for a more general case of unequal damping rates of the atoms $(\gamma_{\text{a1}}\neq \gamma_\text{a2})$ and of the cavity modes $(\gamma_{\text{c1}}\neq\gamma_\text{c2})$. In this case, the spectrum comprises four peaks corresponding to
the transition rates of the four eigenstates $\ket{\epsilon,\varepsilon}$. Two of the peaks are high
(corresponding to the superradiant states~$\ket{-,\varepsilon}$),
and two are short (corresponding to the subradiant states~$\ket{+,\varepsilon}$).
According to our predictions, whenever the high
(short) peaks have the same height and are located in opposite sides
around the cavity frequency, this condition corresponds to a signature of
the underlying maximal entanglement of the superradiant (subradiant) states.
The joint $\gamma_{a1}=\gamma_{a2}$ and $\gamma_{c1}=\gamma_{c2}$
case is special because, in this case, the subradiant
states are optically inactive and only the peaks corresponding
to the superradiant states are present in the spectrum.
Therefore, the symmetry of the spectrum around the cavity frequency increases as
the superradiant states $\ket{-,\varepsilon}$ become increasingly entangled.
Note that, in order to resolve the peaks, the linewidth of
the probe beam has to be sufficiently narrow compared with the cavity and
atomic damping rates.

\subsection{Atoms damped by independent reservoirs}
\label{subsec:atomsdamped}

\begin{figure}[ht]
       \includegraphics[width=4.5cm]{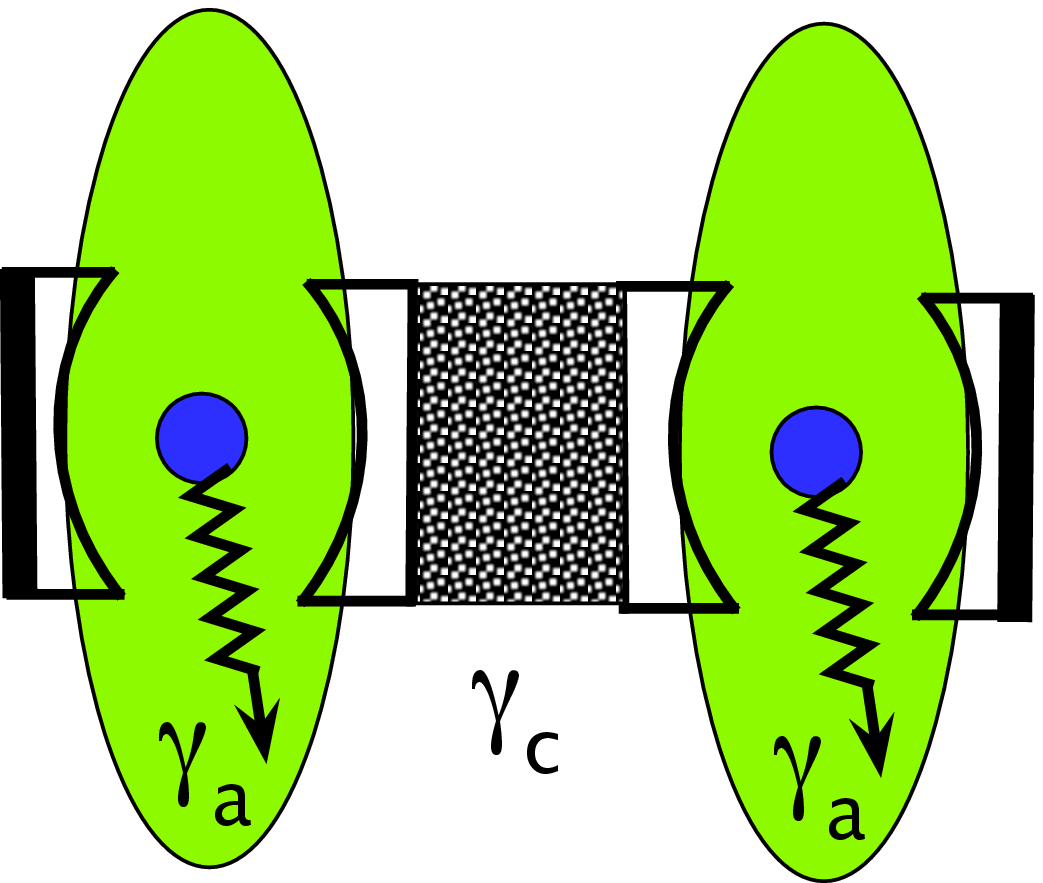}
\caption{(Color online) Two distant cells with the cavity modes overlapping in the shadow area. This area can also be treated as a common reservoir for the cavity modes to which the modes are damped with rate $\gamma_{c}$. The atoms located in the distance cavities are damped by independent reservoirs (green areas) with rate $\gamma_{a}$.}
\label{fig:6}
\end{figure}

In the above analysis we have assumed that the atoms and the cavity modes are damped by common reservoirs.
This assumption is justified if the cells are quite close to each other where the spatial variation of the field can be ignored.
This assumption may not always be true in practice as the cells could be separated by a large distance. In this
case the assumptions that the atoms are coupled to the same reservoir might not be true. For this reason, we
consider now a situation illustrated in Fig.~\ref{fig:6}, in which atoms located in distant cells are damped by independent reservoirs.

In the case of independent reservoirs for the atoms, the transition probabilities from the single excitation states $\ket{\epsilon,\varepsilon}$ to the ground state~$\ket{0}$ are given by
\begin{align}
\Gamma_{\epsilon,\varepsilon} &= \gamma_\text{a}\left(|\bra{\epsilon,\varepsilon}\hat{\sigma}_{1}^{+}\ket{0}|^{2}
+|\bra{\epsilon,\varepsilon}\hat{\sigma}_2^{+}\!\ket{0}\!|^{2}\right) \nonumber\\
&+\gamma_\text{c}|\!\bra{\epsilon,\varepsilon} (a_{1}^{\dag} + a_2^{\dag})\ket{0}|^2 ,
\label{eq14a}
\end{align}
which yields
\begin{equation}
    \Gamma_{\epsilon,\varepsilon} = 2\gamma_\text{a}\left|w_{\epsilon,\varepsilon}\right|^2 +\left(1-\epsilon\right)^2\gamma_\text{c}\left|u_{\epsilon,\varepsilon}\right|^2 .
\label{eq15a}
\end{equation}
We see that, in the case of independent reservoirs for the atoms,
the transition probabilities $\Gamma_{\epsilon,\varepsilon}$ all become different from zero.
There are no trapping (subradiant) states that would form a subspace of decoherence-free states.
However, in the limit of $|w_{\epsilon,\varepsilon}|^2=0$, the states $\epsilon =+$ would form
a decoherence free subspace.

A close look at Eq.~(\ref{uw}) reveals that the amplitudes $|w_{\epsilon,-}|^2$ vanish in the
limit of $r_{\epsilon}\gg 1$.
The condition of $r_{\epsilon}\gg 1$ is met when $(\Delta +\epsilon\kappa)\gg g$ which shows that either the atomic transition frequencies differ significantly from the cavity frequency, $\Delta \gg 0$, or the cells are strongly coupled to each other, $\kappa \gg0$.
Under this condition,
\begin{equation}
    |w_{\epsilon,-}|^{2} = |u_{\epsilon,+}|^2 \approx 0 ,
\end{equation}
and
\begin{equation}
    |w_{\epsilon,+}|^2 =|u_{\epsilon,-}|^2\approx \frac{1}{2}.
\end{equation}
As a consequence the states of the system in Eq.~(\ref{eq:states}) reduce~to
\begin{align}
    \ket{\epsilon,+} &= \frac{1}{\sqrt{2}}\left(\ket{eg} -\epsilon\ket{ge}\right)\ket{00} ,\nonumber\\
    \ket{\epsilon,-} &= \frac{1}{\sqrt{2}}\left(\ket{10} -\epsilon\ket{01}\right)\ket{gg} ,
\label{istates}
\end{align}
with the corresponding transition probabilities
\begin{equation}
    \Gamma_{\epsilon,+} = \gamma_\text{a} ,\quad
     \Gamma_{\epsilon,-} = \frac{1}{2}\left(1-\epsilon\right)^2\gamma_\text{c} .
\end{equation}
Clearly, the transition probability $\Gamma_{+,-}=0$, so that the state~$\ket{+,-}$ is a decoherence free state.

We stress that, in contrast to the states (\ref{eq:states}) damped by a common reservoir,
the states (\ref{istates}) are product states with atom-cavity decoupling due to
having independent reservoirs and strong inter-cell coupling.
Effectively, the system behaves as a two-qubit system.
Our prediction agrees with the results of de Ponte {\it et al.}~\cite{PMM07}, who considered a system of coupled resonators interacting with independent reservoirs and found that the resonators must be strongly coupled to each other for the accomplishment of the condition leading to the decoherence free states.

\section{Generalization to an arbitrary number of coupled JC cells}\label{sec4}

Although the focus of this work is on two coupled JC cells, we can solve the problem more generally.
In this section we show how to solve elegantly the case of $N$ mutually connected~JC cells with the
Hamiltonian
\begin{equation}
    \hat{H}_{N} = \sum_{i=1}^{N}\hat{H}^{i}_{\text{JC}} -\kappa\sum_{i\neq j}\left(\hat{a}_i^\dagger\hat{a}_{j}
    +\text{H.c.}\right) .\label{eq28u}
\end{equation}
The Hamiltonian describes a system composed of $N$ identically coupled cells.
While this description does not include such features as the direct coupling between the atoms and the spatial
variation of the field modes, it is nevertheless of interest, as the simplest model of a group of collective behaving cells. This case is somewhat artificial for large~$N$ but does yield an instructive generalization of the formalism in addition to providing some connection with condensed-matter studies of coupled JC systems~\cite{HBP06}.

\subsection{Energy spectrum}

Restricting to the single-excitation, $\nu=1$, the space of the system is spanned by $2N$ product states
\begin{align}
    \Big\{&\ket{0g0g\cdots 0e},\ket{0g0g\cdots 1g},\ldots,\nonumber\\
        &\ket{0e0g\cdots 0g},\ket{1g0g\cdots 0g}\Big\} .
\end{align}
In the basis of these states the Hamiltonian (\ref{eq28u}) can be represented by a $2N\times 2N$ matrix. Despite a large size, the matrix can be directly diagonalized and results in $N$-cell superposition states (eigenstates). We find that among the $2N$ states, one can distinguish two sets of $N-1$ degenerate antisymmetric eigenstates. These are
\begin{align}
    \ket{\phi_1} &= \frac{1}{{\cal N}_1}\Big[p_+\ket{0g0g\cdots 0e}-\ket{0g0g\cdots 1g} \nonumber\\
        &+\frac{1}{p_-}\ket{0e0g\cdots 0g}+\ket{1g0g\cdots 0g}\Big] ,\nonumber\\
    \ket{\phi_2}&= \frac{1}{{\cal N}_2}\Big[p_+\Big(\ket{0g0g\cdots 0e}-\ket{0g\cdots 0e0g}\Big)\nonumber\\
        &-\ket{0g0g\cdots 1g}+\ket{0g\cdots 1g0g}\Big],\nonumber\\
        &\vdots \nonumber\\
    \ket{\phi_{N-1}} &= \frac{1}{{\cal N}_2}\Big[p_+\Big(\ket{0g0g\cdots 0e}-\ket{0g0e\ldots 0g}\Big)\nonumber\\
        &-\ket{0g0g\cdots 1g}+\ket{0g1g\cdots 0g}\Big] ,\label{eq28}
\end{align}
with corresponding eigenenergies
\begin{align}
    \omega_1 &= \cdots=\omega_{N-1}\nonumber\\
        &= \omega_{c}-\frac{1}{2}\left(\Delta-\kappa\right)+\sqrt{\frac{1}{4}\left(\Delta-\kappa\right)^2+g^2} ,\label{eq36a}
\end{align}
where
\begin{align}
    {\cal N}_1 &= \sqrt{p_+^2+\frac{1}{p_-^2}+2},\quad {\cal N}_2=\sqrt{2p_+^2+2},\nonumber\\
    p_\pm &= -r\pm\sqrt{r^2+1},\quad r = \frac{\Delta-\kappa}{2g}.
\end{align}

The other set of $N-1$ degenerate antisymmetric eigenstates~is
\begin{align}
    \ket{\phi_N}
        &= \frac{1}{{\cal N}_3}\Big[p_{-}\ket{0g0g\cdots 0e}-\ket{0g0g\cdots 1g}\nonumber\\
            &+\frac{1}{p_+}\ket{0e0g\cdots 0g}+\ket{1g0g\cdots 0g}\Big] ,\nonumber\\
    \ket{\phi_{N+1}}
        &= \frac{1}{{\cal N}_4}\Big[p_-\left(\ket{0g0g\cdots 0e}-\ket{0g\cdots 0e0g}\right)\nonumber\\
            &-\ket{0g0g\cdots 1g}+\ket{0g\cdots 1g0g}\Big] ,\nonumber\\
        &\vdots\nonumber\\
    \ket{\phi_{2N-2}}
        &= \frac{1}{{\cal N}_4}\Big[p_-\left(\ket{0g0g\cdots 0e}-\ket{0g0e\cdots 0g}\right)\nonumber\\
            &-\ket{0g0g\cdots 1g}+\ket{0g1g\cdots 0g}\Big] ,\label{eq31}
\end{align}
with eigenenergies
\begin{align}
    \omega_N &=  \cdots=\omega_{2N-2}\nonumber\\
        &= \omega_{c}-\frac{1}{2}\left(\Delta-\kappa\right)
            -\sqrt{\frac{1}{4}\left(\Delta-\kappa\right)^2 +g^2} ,\label{eq36b}
\end{align}
where
\begin{align}
{\cal N}_3 &= \sqrt{p_-^2+\frac{1}{p_+^2}+2} ,\quad {\cal N}_4=\sqrt{2p_-^2+2} .
\end{align}

The remaining two eigenstates of the system are fully symmetric superposition states
\begin{align}
    \ket{\phi_\pm} &= \frac{1}{{\cal N}_\pm}\Big[\frac{1}{p_\pm'}\Big(\ket{0g0g\cdots 0e}
    +\ket{0g0g\cdots 0e0g}+\cdots \nonumber\\
        &+\ket{0e0g\cdots 0g}\Big)+\ket{0g0g\cdots 1g}+\ket{0g0g\cdots 1g0g} \nonumber\\
        &+\cdots+\ket{1g0g\cdots 0g}\Big] ,\label{eq32}
\end{align}
with eigenenergies
\begin{align}
    \omega_{\pm} = \omega_{\text{c}} &-\frac{1}{2}\left[\Delta+\left(N-1\right)\kappa\right] \nonumber\\
        &\pm\sqrt{\frac{1}{4}\left[\Delta+\left(N-1\right)\kappa\right]^2 +g^2} ,\label{eq36c}
\end{align}
respectively, where
\begin{equation}
        {\cal N}_\pm=\sqrt{\frac{N}{p_\pm'^2}+N} ,
\end{equation}
and
\begin{align}
    p'_\pm = -r'\pm\sqrt{r'^2+1} ,\quad
    r' = \frac{\Delta+\left(N-1\right)\kappa}{2g} .
\end{align}

We note from Eqs.~(\ref{eq36a}), (\ref{eq36b}) and (\ref{eq36c}) that for $g=0$, energies of the antisymmetric states cross at $\Delta =\kappa$,
whereas energies of the two fully symmetric states cross at $\Delta=-(N-1)\kappa$. When $g\neq 0$,
avoiding crossings occur at $\Delta=\kappa$ for the antisymmetric states, and at $\Delta = -(N-1)\kappa$ for the symmetric states. If and only if $\Delta = -(N-1)\kappa$ is true, do the eigenstates $\ket{\phi_\pm}$ become W-type maximally entangled states.
\begin{figure}[ht]
    \includegraphics[width=4.1cm]{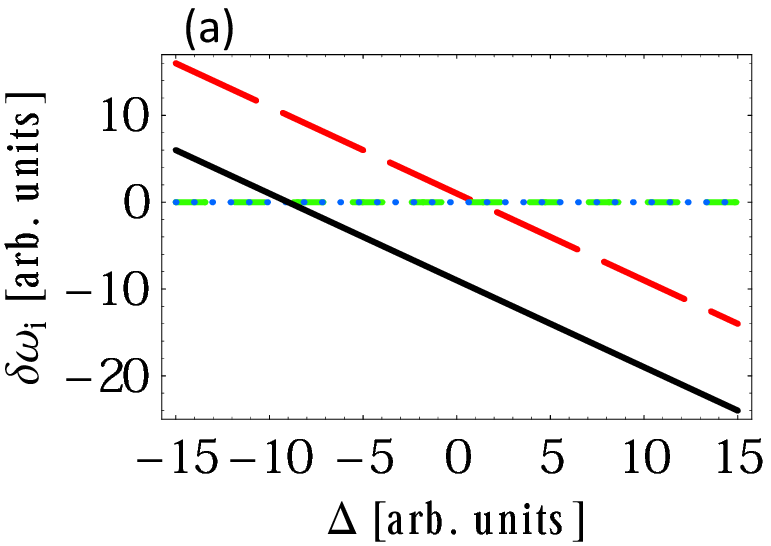}
    \includegraphics[width=4.1cm]{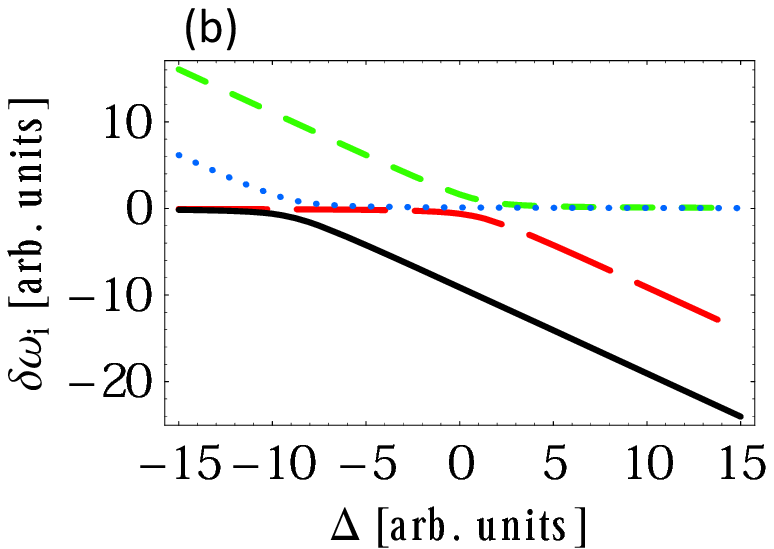}
    \caption{(Color online) The dependence of the energy difference
    $\delta\omega_{i} = \omega_{i}-\omega_{\text{c}}$ on the detuning $\Delta$ (in arbitrary units) of
    the $\nu=1$ eigenstates of $N=10$ coupled JC cells for $\kappa=1$ and (a)~$g =0$, (b)~$g =1$. The green short dashed line is for the set of antisymmetric states $\ket{\phi_{i}}\, (i=1,2,\ldots,N-1)$, the red long dashed line is for the set of antisymmetric states $\ket{\phi_{j}}\, (j=N, N+1,\ldots,2N-2)$, and the blue dotted and black solid lines are for the symmetric states $\ket{\phi_{+}}$ and $\ket{\phi_{-}}$, respectively.}
\label{fig5}
\end{figure}

Figure~\ref{fig5} shows the dependence of the energies of the eigenstates $\ket{\phi_\pm}$ on the detuning $\Delta$. We see two crossing points when $g=0$ and the avoided crossing between the states when $g\neq 0$. It clearly illustrates the appearance of two separate groups of degenerate antisymmetric states with the minimum energy separation $2g$ at $\Delta = \kappa$, and two symmetric states with the minimum energy separation $2g$ at $\Delta =-(N-1)\kappa$. Perhaps the most interesting aspects of the $N$ cell system is
that the energies of the antisymmetric states are independent of $N$.
Adding more cells increases the number of the
antisymmetric states but does not affect their energies.
Here the antisymmetry is with respect to a permutation of the two entangled JC
cells, not of any pair of `qubits', and it is exact only in the limit of maximal entanglement $|p_{\pm}|=1$.
For the specific case $N=2$, the antisymmetric states
$|\phi_{1,2}\rangle$ correspond to the (subradiant) states
$|+,i\rangle$ introduced in Eq.~(\ref{eq:states}), whereas the
symmetric states~$\ket{\phi_{\pm}}$ correspond to the superradiant
ones $\ket{-,i}$. Note also that $N=2$ is the only case in which
the antisymmetric and symmetric states have entanglement between the
same number of `qubits', four in this case.

\subsection{Transition rates}

Having derived the explicit forms of the eigenstates $\ket{\phi_{i}}$ of the single-excitation sector $\nu=1$, we now turn to calculate transition rates between the eigenstates $\ket{\phi_i}$ and the ground state $\ket{0}=\ket{0g\cdots 0g}$. We consider separately two cases. In the first, we assume that the atoms are coupled to a common reservoir. In the other case, we assume that the atoms are coupled to independent reservoirs.

When the atoms are coupled to a common reservoir, the transitions occur with rates
\begin{equation}
    \Gamma_i=\left|\sum_{j=1}^N\sqrt{\gamma_{\text{a}j}}\bra{\phi_i}\hat{\sigma}_j^\dagger\ket{0}\right|^2
        +\left|\sum_{j=1}^N\sqrt{\gamma_{\text{c}j}}\bra{\phi_i}\hat{a}_j^\dagger\ket{0}\right|^2 .
\label{eq:trans}
\end{equation}

Applying Eqs.~(\ref{eq28}), (\ref{eq31}) and (\ref{eq32}), after straightforward calculations we obtain
\begin{align}
\Gamma_1 &=\frac{1}{\mathcal{N}_1^2}\left[\left(p_+\sqrt{\gamma_{\text{a}1}}+\frac{\sqrt{\gamma_{\text{a}N}}}{p_-}\right)^2
+\left(\sqrt{\gamma_{\text{c}N}}-\sqrt{\gamma}_{\text{c}1}\right)^2\right] ,\nonumber\\
\Gamma_i &= \frac{1}{\mathcal{N}_2^2}\left[p_+^2 \left(\sqrt{\gamma_{\text{a}1}}-\sqrt{\gamma_{\text{a}i}}\right)^2
+\left(\sqrt{\gamma_{\text{c}i}}-\sqrt{\gamma}_{\text{c}1}\right)^2\right] , \nonumber\\
i&=2,\ldots,N-1, \nonumber\\
\Gamma_N &=\frac{1}{\mathcal{N}_3^2}\left[\left(p_-\sqrt{\gamma_{\text{a}1}}+\frac{\sqrt{\gamma_{\text{a}N}}}{p_+}\right)^2
+\left(\sqrt{\gamma_{\text{c}N}}-\sqrt{\gamma}_{\text{c}1}\right)^2\right] ,\nonumber\\
\Gamma_j &=\frac{1}{\mathcal{N}_4^2}\left[p_-^2\left(\sqrt{\gamma_{\text{a}1}}
            -\sqrt{\gamma_{\text{a}j}}\right)^2
        +\left(\sqrt{\gamma_{\text{c}j}}
            -\sqrt{\gamma}_{\text{c}1}\right)^2\right] ,\nonumber\\
      j&=N+1,\ldots,2N-2, \nonumber\\
\Gamma_{\pm} &=\frac{1}{\mathcal{N}^2_\pm}\left[\frac{1}{p^{'2}_{\pm}}\left(\sum_{i=1}^N\sqrt{\gamma_{\text{a}i}}\right)^{2} +\left(\sum_{i=1}^N\sqrt{\gamma_{\text{c}i}}\right)^{2}\right] .\label{eq46}
\end{align}

For the case of identical cells,
\begin{equation}
    \gamma_{\text{a}1}=\cdots=\gamma_{\text{a}N}\equiv \gamma_{\text{a}}
\end{equation}
and
\begin{equation}
    \gamma_{\text{c}1}=\cdots=\gamma_{\text{c}N}\equiv \gamma_{\text{c}},
\end{equation}
it follows that the transition rates from all of the antisymmetric states are zero
\begin{equation}
    \Gamma_1=\cdots=\Gamma_{N}=\cdots=\Gamma_{2N-2}=0 .\label{eq47}
\end{equation}
Only the fully symmetric states $\ket{\phi_\pm}$ are optically active and are damped at rates
\begin{align}
\Gamma_{\pm} = \frac{N}{\left(1+p^{\prime 2}_{\pm}\right)}\left(\gamma_{\text{a}} +p^{\prime 2}_{\pm}\gamma_{\text{c}}\right) ,\label{eq48}
\end{align}
in which we see the characteristic $N$ time enhancement, the superradiant behaviour~\cite{dic,eb71}.

From Eqs.~(\ref{eq47}) and (\ref{eq48}) it is apparent that, just as in the $N=2$ case, the $2N-2$ antisymmetric states are all optically inactive, whereas the two symmetric ones are superradiant.
The distinction is quite convenient because higher-dimensional entanglement is present in the latter states,
so a symmetric absorption spectrum would reveal maximal $2N$-partite W-like entanglement shared among all the JC cells.

\begin{figure}[ht]
       \includegraphics[width=6.3cm]{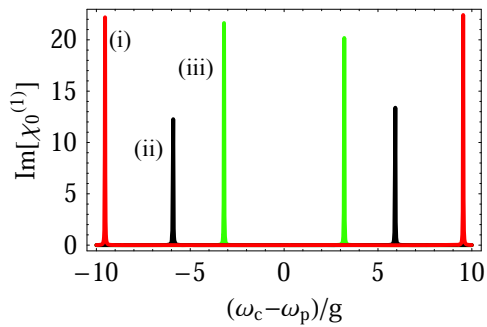}
    \caption{(Color online) The absorption spectra ${\rm Im}\left[\chi^{(1)}_{0}(\omega_\text{p})\right]$
    of coupled $N=10$ cells plotted as a function of $(\omega_\text{c}-\omega_\text{p})/g$
    for $\gamma_\text{a}/g=0.05$,  $\gamma_\text{c}/g=0.02$, $\kappa/g=2$, and different detunings
    $\Delta$: $\Delta/g=1$ (green line (iii)), $\Delta/g=10$ (black line (ii)), and $\Delta/g=-18$ (red line (i)).}
\label{fig:ab}
\end{figure}

As the $N$-cell system radiates only from the fully symmetric states, the Hilbert space
of the system is confined to the totally symmetric subspace of two states $\ket{\phi_\pm}$.
Given that there are only two non-zero transition rates from the excited states to the ground state,
the absorption spectrum of a weak probe beam monitoring the system is expected to be composed of two peaks.
This feature is seen in Fig.~\ref{fig:ab}, which displays the absorption spectrum for coupled $N=10$ cells.

As expected, there are two peaks corresponding to transitions from the states $\ket{\phi_\pm}$ to the ground state $\ket 0$. The relative peak amplitudes are a function of $\Delta$, and if and only if $\Delta =-(N-1)\kappa$, the amplitudes are equal and symmetrically located about
\begin{equation}
    \omega_\text{c}-\omega_p=\pm g .
\end{equation}
When $\Delta =-(N-1)\kappa$, the states $\ket{\phi_{\pm}}$ reduce to maximally entangled W states.
Hence, a symmetric absorption spectrum is a signature of a maximally-entangled W states with $2N$ degrees of freedom.

We close this section by evaluating the transition rates for the case when the atoms are damped by independent reservoirs. The total transition rate due to the interaction of the atoms with independent reservoirs and the cavity modes damped by a common reservoir is defined by
\begin{equation}
    \Gamma_i=\sum_{j=1}^N\gamma_{\text{a}j}\left|\bra{\phi_i}\hat{\sigma}_j^\dagger\ket{0}\right|^2
        +\left|\sum_{j=1}^N\sqrt{\gamma_{\text{c}j}}\bra{\phi_i}
            \hat{a}_j^\dagger\ket{0}\right|^2 .\label{eq:transi}
\end{equation}
When we make use  of Eqs.~(\ref{eq28}), (\ref{eq31}) and (\ref{eq32}) in Eq.~(\ref{eq:transi}), we readily obtain
\begin{align}
\Gamma_1 &=\frac{1}{\mathcal{N}_1^2}\left[\left(p_+^2\gamma_{\text{a}1}+\frac{\gamma_{\text{a}N}}{p_-^2}\right)
+\left(\sqrt{\gamma_{\text{c}N}}-\sqrt{\gamma}_{\text{c}1}\right)^2\right] ,\nonumber\\
\Gamma_i &=\frac{1}{\mathcal{N}_2^2}\left[p_+^2\left(\gamma_{\text{a}1}+\gamma_{\text{a}i}\right)
+\left(\sqrt{\gamma_{\text{c}i}}-\sqrt{\gamma}_{\text{c}1}\right)^2\right] ,\nonumber\\
i &=2,\ldots,N-1 ,\nonumber\\
\Gamma_N &=\frac{1}{\mathcal{N}_3^2}\left[\left(p_-^2\gamma_{\text{a}1}+\frac{\gamma_{\text{a}N}}{p_+^2}\right)
+\left(\sqrt{\gamma_{\text{c}N}}-\sqrt{\gamma}_{\text{c}1}\right)^2\right] ,\nonumber\\
\Gamma_j &=\frac{1}{\mathcal{N}_4^2}\left[p_-^2\left(\gamma_{\text{a}1} +\gamma_{\text{a}j}\right)
        +\left(\sqrt{\gamma_{\text{c}j}} -\sqrt{\gamma}_{\text{c}1}\right)^2\right] ,\nonumber\\
j&=N+1,\ldots,2N-2 ,\nonumber\\
\Gamma_{\pm} &=\frac{1}{\mathcal{N}^2_\pm}\left[\frac{1}{p^{'2}_{\pm}}\sum_{i=1}^N\gamma_{\text{a}i}+\left(\sum_{i=1}^N\sqrt{\gamma_{\text{c}i}}\right)^2\right] .\label{eq55}
\end{align}
It is easy to see that the transition rates~(\ref{eq55}) differ significantly from that we encountered in Eqs.~(\ref{eq46}) for the decay of the atoms in a common reservoir. Notice that all of the transition rates are now different from zero, and none of the rates can be reduced to zero, indicating no subradiance. This principle also applies to the special case of identical cells,
\begin{equation}
    \gamma_{\text{a}1}=\cdots=\gamma_{\text{a}N}\equiv \gamma_{\text{a}}
\end{equation}
and
\begin{equation}
    \gamma_{\text{c}1}=\cdots=\gamma_{\text{c}N}\equiv \gamma_{\text{c}}.
\end{equation}

However, despite the fact that the transition rates (\ref{eq55}) are different from zero, we find that in the case of identical cells and under a strong coupling between the cells $(\kappa\gg g)$, the transition rates reduce to
\begin{align}
\Gamma_{1} &\approx \gamma_{\text{a}} ,\nonumber\\
\Gamma_{i} &\approx 0 ,\quad i=2,\ldots,N-1 ,\nonumber\\
\Gamma_{j} &\approx \gamma_{\text{a}} ,\quad j=N,\ldots,2N-2 ,\nonumber\\
\Gamma_{\pm} &\approx  \frac{1}{\left(1+p^{\prime 2}_{\pm}\right)}\left(\gamma_{\text{a}} +p^{\prime 2}_{\pm}N\gamma_{\text{c}}\right) .
\end{align}
Clearly, in the limit of a strong coupling the states $\ket{\phi_{i}}\, (i=2,\ldots,N-1)$ do not decay, so they form a decoherence-free subspace. The remaining states decay with nonzero rates. The states $\ket{\phi_{1}}, \ket{\phi_{N}}$ and $\ket{\phi_{j}}$ decay with the rate equal to the single atom decay rate $\gamma_{\text{a}}$, whereas the rates $\Gamma_{\pm}$ show the characteristic $N$ times enhancement (superradiance), but only with respect to the cavity damping. The contribution to $\Gamma_{\pm}$ from the atomic part, proportional to $\gamma_{\text{a}}$, is independent of $N$. This is what one could expect, since the atoms are damped by independent reservoirs. This is also consistent with the results of de Ponte {\it et al.}~\cite{PMM07}.

\section{Conclusions}\label{sec5}

We have studied a system of coupled Jaynes-Cummings~(JC) cells with
coupling created by overlapping evanescent cavity fields or by
tunneling of photons between the cells. This overlap or tunneling
determines the coherent hopping rate between the pair of cavities.
This coherent hopping term along with the strength of atom-field
coupling within each cavity determines the unitary dynamics of the
closed coupled system. Our analysis also includes incoherent cavity
damping and atomic spontaneous emission rates. These incoherent
processes are unavoidable in experiments but are sometimes not
considered in studies of condensed-matter types of properties of
such systems. We included these incoherent terms for completeness
and found that ensuring differences in incoherent rates can be quite
valuable in detecting entanglement of these coupled atom-cavity systems.

The principal advantages of our treatment are that it allows to study
the relation between the radiative properties of the system
and entanglement. For example, in the case of the single JC cell we have observed
that the ratio of the transition rates for the
two lowest-doublet states directly reveal the degree of entanglement between the atom
and cavity mode provided that the spontaneous emission
rate~$\gamma_\text{a}$ is not equal to the cavity loss
rate~$\gamma_\text{c}$. The degree of entanglement depends on the
values of atom-cavity coupling strength~$g$ in the JC Hamiltonian.

We have also studied transition rates from the second doublet to the first doublet for the single JC model.
In this case we have transitions from entangled $n=2$ state to entangled $n=1$ states,
in contrast to the $n=1$ to $n=0$ transition from an entangled first-doublet state to a ground state,
which is an atom-cavity product state. We have found that the rates depend explicitly on properties of states in both doublets, namely that it depends explicitly on both $\theta_n$ and $\theta_{n-1}$,
but adding the transitions rates together as in Eq.~(\ref{eq:Gamma_+n}) only depends on~$\theta_n$.
Thus the transition rates directly reveal the degree of entanglement of the $n^\text{th}$ doublet without being mixed
up with the degree of entanglement in the lower doublet.
Direct measurement of entanglement of the states is possible.

These results on entanglement in the single JC cell are interesting and provide a foundation to study coupled JC cells.
For coupled JC cells we focus only on the $\nu=1$ case. We have derived the explicit analytical forms of
the transition rates between the $\nu=1$ quadruplet of states for the double JC cell to the (product) ground state.
There are four rates in this case, and each of these rates conveys information about the degree of entanglement in the state including whether the states are maximally entangled.
When loss rates from cells (both spontaneous emission and cavity loss) are equal,
information about the degree of entanglement is not obtained by our method but is obtained for unequal rates.

Finally, we have shown how to extend our two JC cell case to an arbitrary number $N$ of the JC cells
such that each cell is coupled to all other cells. We have established
conditions for maximal entanglement and calculate the
susceptibility. This $N$ cell case is likely only of theoretical
interest as it is hard to imagine how such a system would be
configured in a real experiment, but this general extension to~$N$
cells is interesting nonetheless.

\acknowledgments

This work has been supported by NSERC, MITACS, CIFAR, QuantumWorks,
iCORE, China's 973 Program 2011CB911203 and NSFC 11004029 and
11174052. BCS is supported by a CIFAR Fellowship.

\end{document}